\documentclass{PoS}

\PoS{PoS(LAT2005)033}

\title{Twisted mass fermions: neutral pion masses from disconnected
contributions}

\ShortTitle{TM disconnected}

\author{
F. Farchioni$^a$, K. Jansen$^b$, C. McNeile$^c$, C. Michael$^c$,
 I. Montvay$^d$, K. Nagai$^b$, M. Papinutto$^b$, \speaker{J. Pickavance}$^c$,
 E. Scholz$^d$, L. Scorzato$^e$, A. Shindler$^b$, N. Ukita$^d$, 
 C. Urbach$^{bf}$, U. Wenger$^b$ and I. Wetzorke$^b$\\
 \llap{$^a$}Institut f\"{u}r Theoretische Physik, Universit\"{a}t M\"{u}nster, 
 Wilhelm-Klemm-Str. 9, 48149 M{\"u}nster, Germany \\
 \llap{$^b$}John von Neumann-Institut f\"ur Computing NIC,
   Platanenallee 6, 15738 Zeuthen, Germany \\ 
 \llap{$^c$}Theoretical Physics Division, Dept of Mathematical Sciences,  
   University of Liverpool, Liverpool L69 3BX, UK\\
 \llap{$^d$}Deutsches Elektronen-Synchrotron DESY, 
 Notkestr.~85, 22603 Hamburg, Germany\\ 
 \llap{$^e$}Institut f\"{u}r Physik, Humboldt Universit\"{a}t zu Berlin,
 Newtonstr.\ 15, 12489 Berlin, Germany \\
 \llap{$^f$}Institut f\"{u}r Theoretische Physik, Freie Universit\"{a}t Berlin,
   Arnimallee 14, 14195 Berlin, Germany \\

\email{${}^*$jennyp@liv.ac.uk} 
}

\abstract{ Twisted mass fermions allow light quarks to be explored but
with  the consequence that there are mass splittings, such as between 
the neutral and charged pion. Using a direct calculation of the
connected neutral pion correlator and stochastic methods to evaluate the
disconnected correlations, we determine the neutral pion mass. We
explore the dependence on  lattice spacing and quark mass in quenched
QCD. For dynamical QCD, we determine the  sign of the splitting which is
linked, via chiral PT, to the nature of the  phase transition at small
quark mass.
 }

\FullConference{XXIII International Symposium on Lattice Field Theory\\  
                25th-30th July 2005\\
	        School of Mathematics, Trinity College, Dublin}

\usepackage{graphicx}

\begin{document}

\section{TMQCD - Introduction}

{
 TMQCD has Wilson quarks with the mass term modified by { $m \to m+i \mu
\tau_3 \gamma_5$}. In the continuum limit,  a chiral rotation shows that
 this is  equivalent to conventional QCD. At non-zero lattice spacing, however, 
the twisted mass formulation has several advantages, especially at 
maximal twist (where the effective mass term is imaginary).
 These advantages include (see \cite{shindler05} for a review)
 \begin{itemize}
  \item improved behaviour of inverters  for small  pion mass, \\
  \item order { $a^2$} discretisation errors of flavour-conserving
quantities, but with, however, 
 the disadvantage that there will be  flavour breaking effects (eg. {
$m(\pi^+) \ne m(\pi^0$)}) at non-zero lattice spacing
 \end{itemize}

Here we explore the $\pi^+$, $\pi^0$ mass difference  both in quenched
QCD to check scaling (with { $a$})  and  in full QCD ($N_f=2$) to
explore the relationship via Chiral PT  models to the nature of the
phase transition at small quark masses.
 }

 {At maximal (90$^0$) twist, the operators appropriate to create hadrons 
are shown below:
 
\begin{tabular}{llllllllll}
 hadron & operator & Disconnected piece \\
 $\pi^+  $& $\overline{u} \gamma_5 d $ &\\
 $\pi^0  $& $\overline{u} I u + \overline{d} I d $ &  { Re Tr $G_u$}\\
 $\eta^0  $& $\overline{u} I u - \overline{d} I d $ & {  $i$Im Tr $G_u$}\\
 $a_0^+    $& $\overline{u} I d $ &\\
 $a_0^0  $& $\overline{u} \gamma_5 u + \overline{d} \gamma_5 d $& { Re Tr $\gamma_5 G_u$}\\
 $f_0   $& $\overline{u} \gamma_5 u - \overline{d} \gamma_5 d $ &{ $i$Im Tr $\gamma_5 G_u$}\\
 \end{tabular}

\medskip

 \noindent where we have made use of the identity for  propagators that
$G_d=\gamma_5 G_u^{\dag} \gamma_5$. In several cases we need
disconnected contributions: the notation  is that $ {\rm Tr} G_u(t)
= \sum_x \overline{u}(x,t) u(x,t)$ and the disconnected contribution  is
the correlation of $G_u(0) G_u(t)$, with the vacuum piece of $G_u$ 
subtracted. 
 This is illustrated by the right hand contribution below: 
}

 \vspace{0.3cm}
\includegraphics[width=7cm]{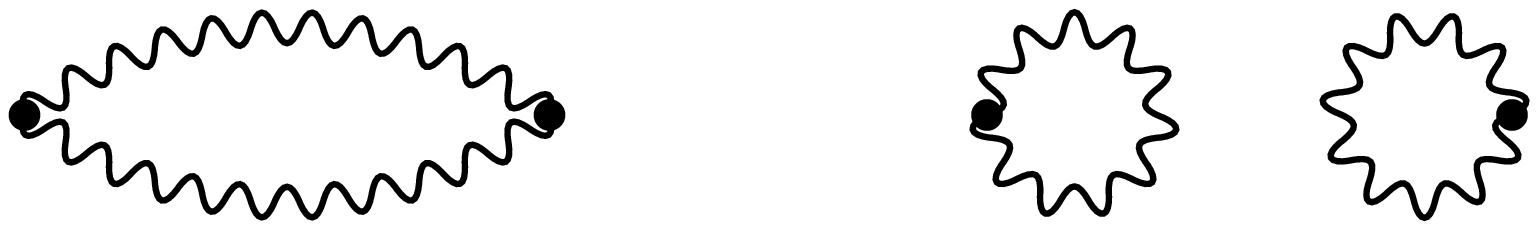}

\section {Lattice evaluation of $\pi^0$ mass}

 To evaluate the disconnected contributions to the hadronic correlation, we
 use  a stochastic volume source with enough samples (24) that 
the stochastic sampling error is not dominant. We
measure all $\overline{q} \Gamma q$ loops at all t (for zero momentum) and 
 use both  local and fuzzed sources. We use variance
reduction~\cite{McNeile:2000xx}  (gaussian
 random noise with subtraction of the lowest-order hopping  parameter
term) and we  check the stochastic method by also evaluating the 
connected $\pi^0$ correlator which we compare with conventional (fixed source) 
evaluation. 

 For quenched lattices at $\beta=6.0$ and $5.85$ with Wilson gauge
action,   $\kappa$ was  chosen~\cite{Jansen:2005cg} to minimise parity
mixing at $\mu=0$.   The quenched results presented here are also
reported in ref~\cite{Jansen:2005cg}.

As shown in fig.~1a, we find that the disconnected contribution is
dominant in the  $\pi^0$ correlator. This is advantageous since the
large signal is more easily evaluated using stochastic methods.

\begin{figure}[thb]
 \begin{center}
 \vspace{6.0cm} 

 \includegraphics{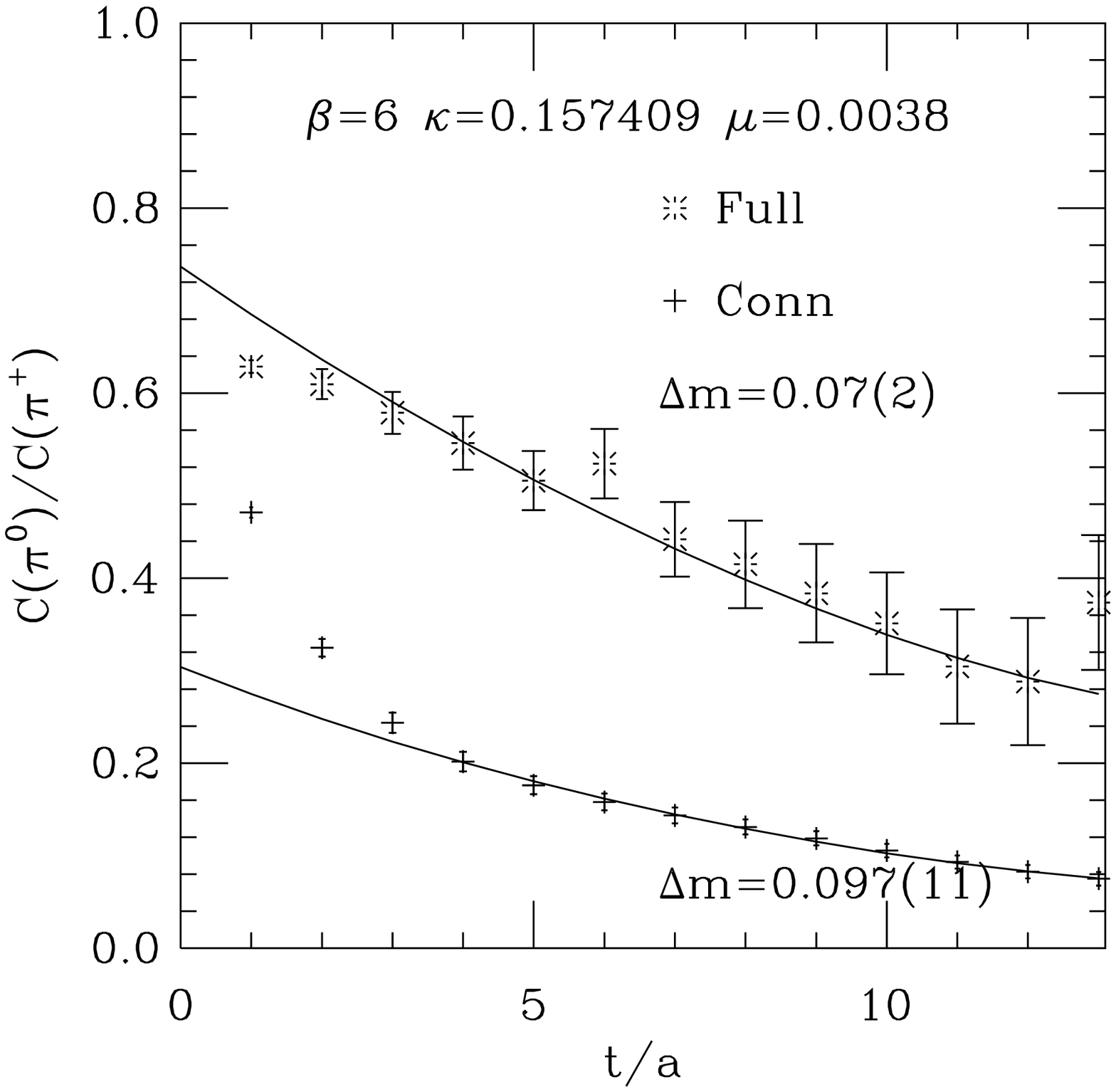}
 \includegraphics{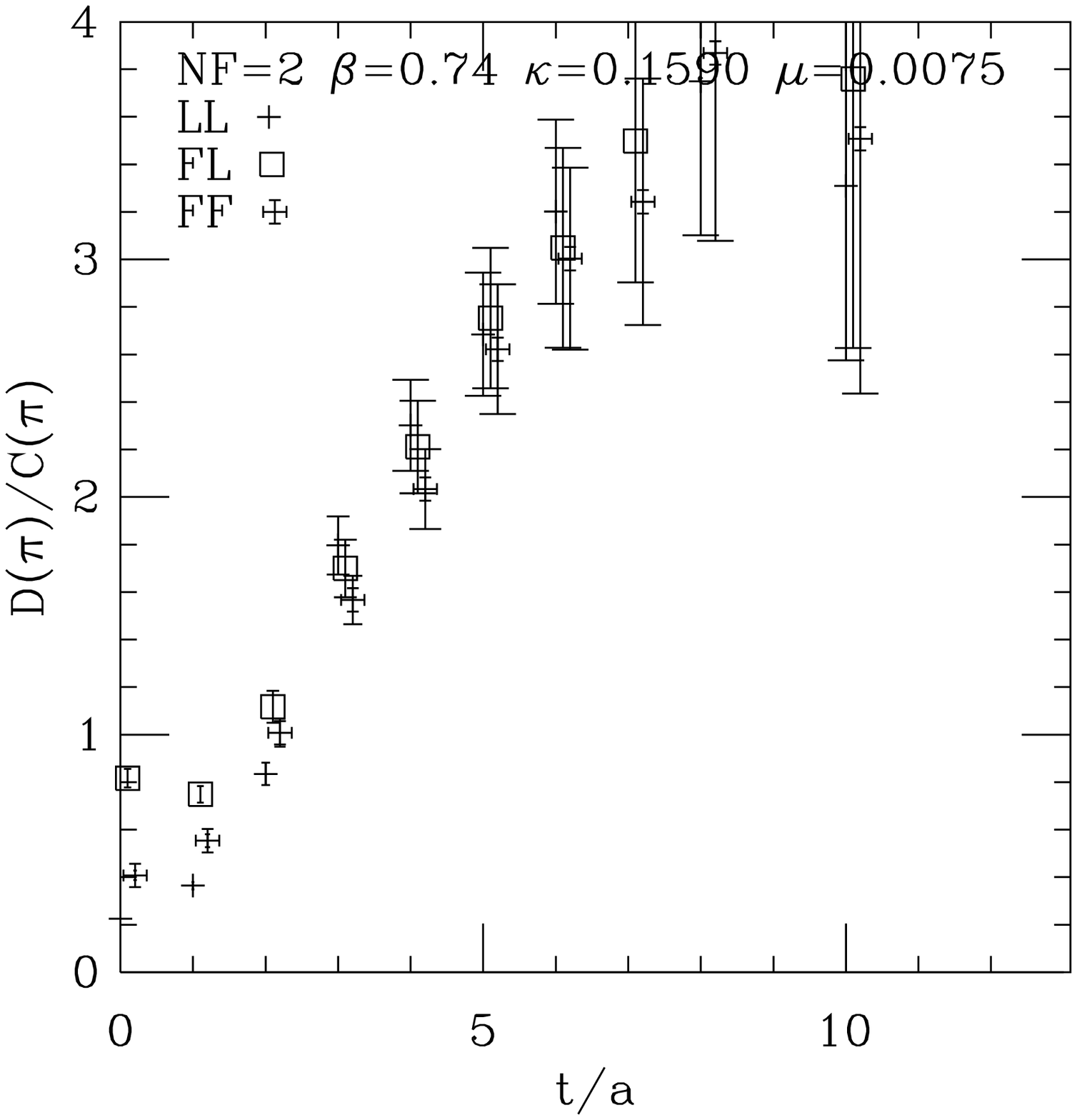}
 \end{center}

 \caption{(a) Ratio of pion correlators (with local operators) for quenched
fermions;
 (b)  Ratio of  disconnected to connected contribution to $\pi^0$ for
dynamical fermions
 }
 \end{figure}

\begin{table}[h]
 \caption{Pion masses (in lattice units)} 
\begin{tabular}{llllllllll}

\hline
N$_f$& N$_g$ &  $\beta$ &  $\kappa$ & a$\mu$ & $ r_0/a$ & $\pi^+$ & 
       $\pi^0$(conn) & $\pi^0$(tot)& $\pi^0-\pi^+$\\
\hline   
0&  400&    6.0   &.157409&.0038& 5.368&.1217(66) &.218(9)  &.19(2)     &.07(2)\\
0&  400&    6.0   &.157409&.0076& 5.368&.1708(50) &.246(5)  &.19(2)     &.02(2)\\
0&  400&    6.0   &.157409&.0109& 5.368&.2047(27) &.267(4)  &.23(2)     &.02(2)\\

0&  100&    5.85  &.162379&.0050& 4.067&.1640(23)  &.319(14) &.28(5)     &.12(5)\\
0&  100&    5.85  &.162379&.0100& 4.067&.2289(17)  &.351(8)  &.25(5)     &.02(5)\\
0&  100&    5.85  &.162379&.0144& 4.067&.2736(14)  &.380(6)  &.28(6)     &.00(6)\\
\hline 
2&  250&    .74   &.1590  &.0075& 3.87 &.1954(22)  &.37(4)   &.24(3)     &.04(3)\\
\hline 
2&  225&    3.9   &.1610  &.0075& 5.28 &.214(7)  &.35(2)   &.27(2)     &.06(2)\\
\hline
 
\end{tabular}

\end{table}

 We also plot the ratio of the full $\pi^0$ correlator to the $\pi^+$
correlator,  and this is not consistent with 1.0. We quantify this
flavour splitting by fitting  both $\pi^0$ and $\pi^+$ correlators
separately (using both local and non-local sources)  and  we show our
results in Table~1.

As well as evaluating the full $\pi_0$ correlator, we can take advantage
of the more accurate  results available for the connected contribution
to the $\pi^0$ by interpreting~\cite{Jansen:2005cg} this  contribution
as from a (quenched) theory with fermion action with mass term  $m+i\mu
\gamma_5$. That theory has  no disconnected contribution to $\pi_0$
(since isospin is now a good symmetry)  but the difference between it
and the TMQCD value for the $\pi^+$ mass can be  used to explore lattice
artifacts. These results are illustrated in fig.~2.  Here we see that 
the $\pi^0$ - $\pi^+$ mass difference decreases consistently with $a^2$
as $a \to 0$. Moreover, we find that we can summarise the splitting by 
  $r_0^2 (m^2(\pi^0)-m^2(\pi^+)) \approx c(a/r_0)^2$, with the   
  difference smaller including the disconnected contribution to the 
$\pi^0$ ($c \approx 10$ but with large error) than connected alone($c
\approx 23$ and quite well determined).

\begin{figure}[htb]
\vspace{6.0cm} 
 \begin{center}

\includegraphics{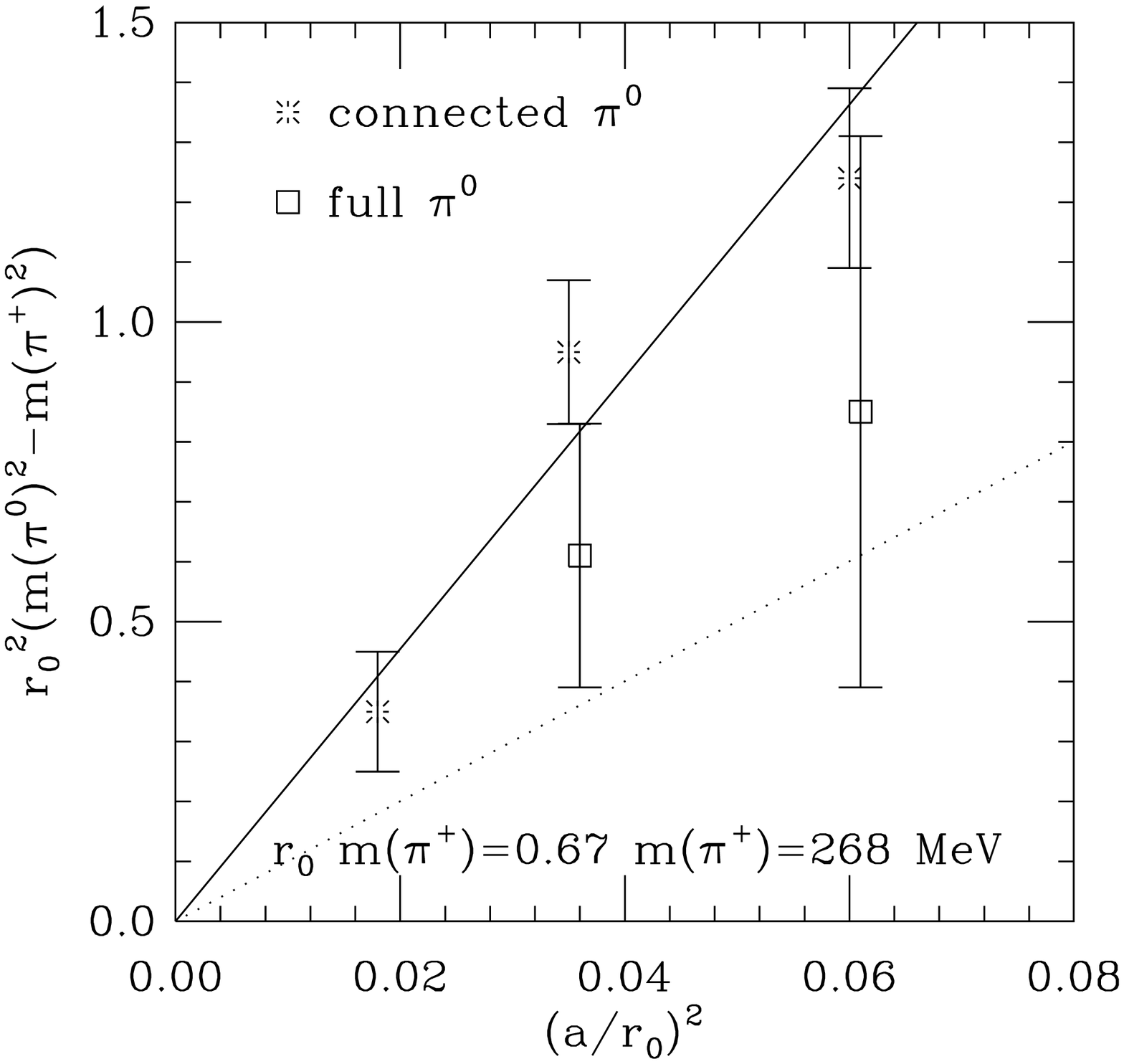}
\includegraphics{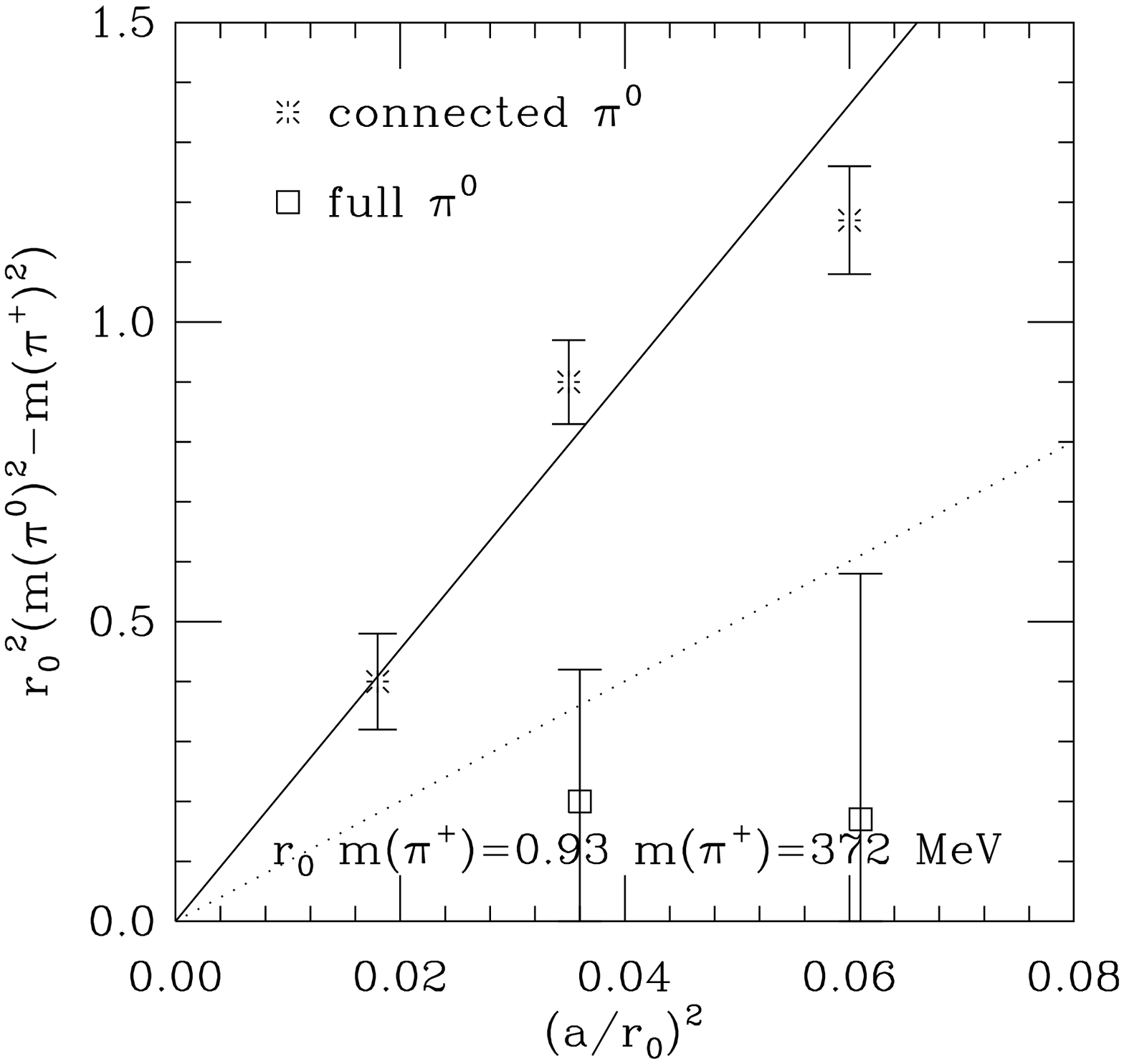}

\end{center}
 \caption{Mass splittings between $\pi^0$ and $\pi^+$ in quenched studies.}


\end{figure}

 We also explore the $\pi^0$ mass with dynamical
fermions~\cite{farchioni}  - using $N_f=2$  flavours of quarks ($u$ and
$d$). We fitted the $2 \times 2$ matrix of local and fuzzed  correlators
where both connected and disconnected contributions were here evaluated
stochastically.  We concentrate on results with approximately  maximal
twist. Using the DBW2 gauge action ($\beta=0.74$) (see also fig.~1b),
the pion masses  are recorded in Table~1.  The mass splitting $m(\pi^0)
- m(\pi^+)$ is consistent with zero but  with a sign which corresponds
(via leading order $\chi$ PT~\cite{Sharpe:2004ny}) to an Aoki phase at
small quark mass. We also analyse results from the tree-level improved
Symanzik  gauge action (TLS $\beta=3.9$) which are also presented  in
Table~1. The case nearest to maximal twist again shows a mass splitting
$m(\pi^0) - m(\pi^+)$   with a sign which corresponds with similar
assumptions to an Aoki phase.

\section {Other mesons}

 The flavour singlet pseudoscalar ($\eta$ for $N_f=2$) involves a disconnected 
contribution even for conventional Wilson fermions. Here we evaluate 
the contribution for our twisted data sets. The full 
contribution is $C-D$ where $C$ is connected and $D$ is disconnected.
 Positivity then requires $C-D > 0$ or, equivalently, $D/C < 1$. 

 We see, in fig.~3a that this requirement is violated in quenched QCD.  
 Quenched QCD has ghost states and the flavour-singlet sector 
is not physical. For $N_f=2$, however, the requirement $D/C < 1$ is 
well satisfied and it would be possible to extract the $\eta$ mass from 
the rate at which it approaches 1, albeit with large errors.  This is confirmation 
that the algorithms used to evaluate dynamical gauge configurations are 
consistent with positivity requirements. 

\begin{figure}[htb]

\vspace{6.5cm} 

\includegraphics{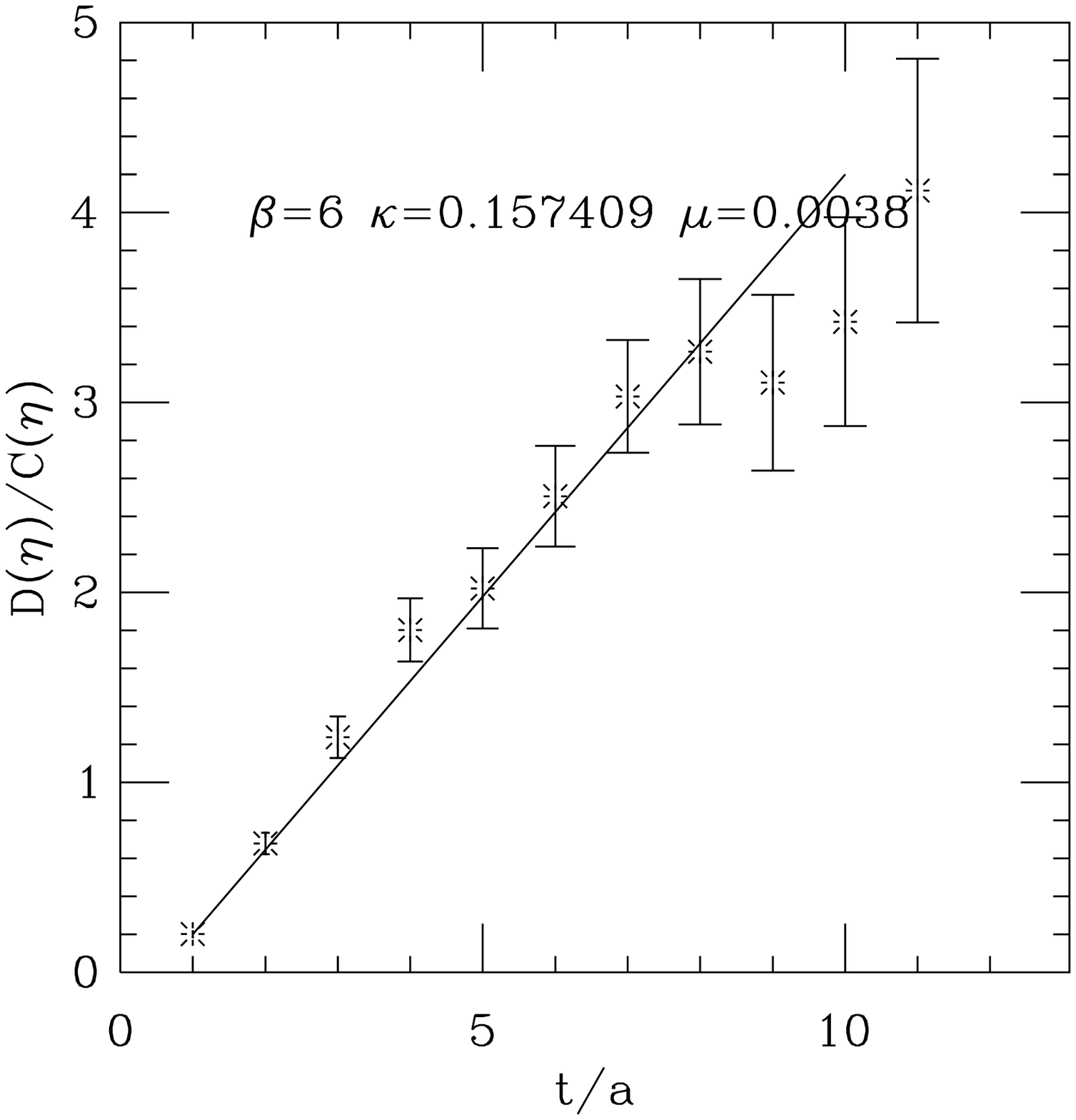}
\includegraphics{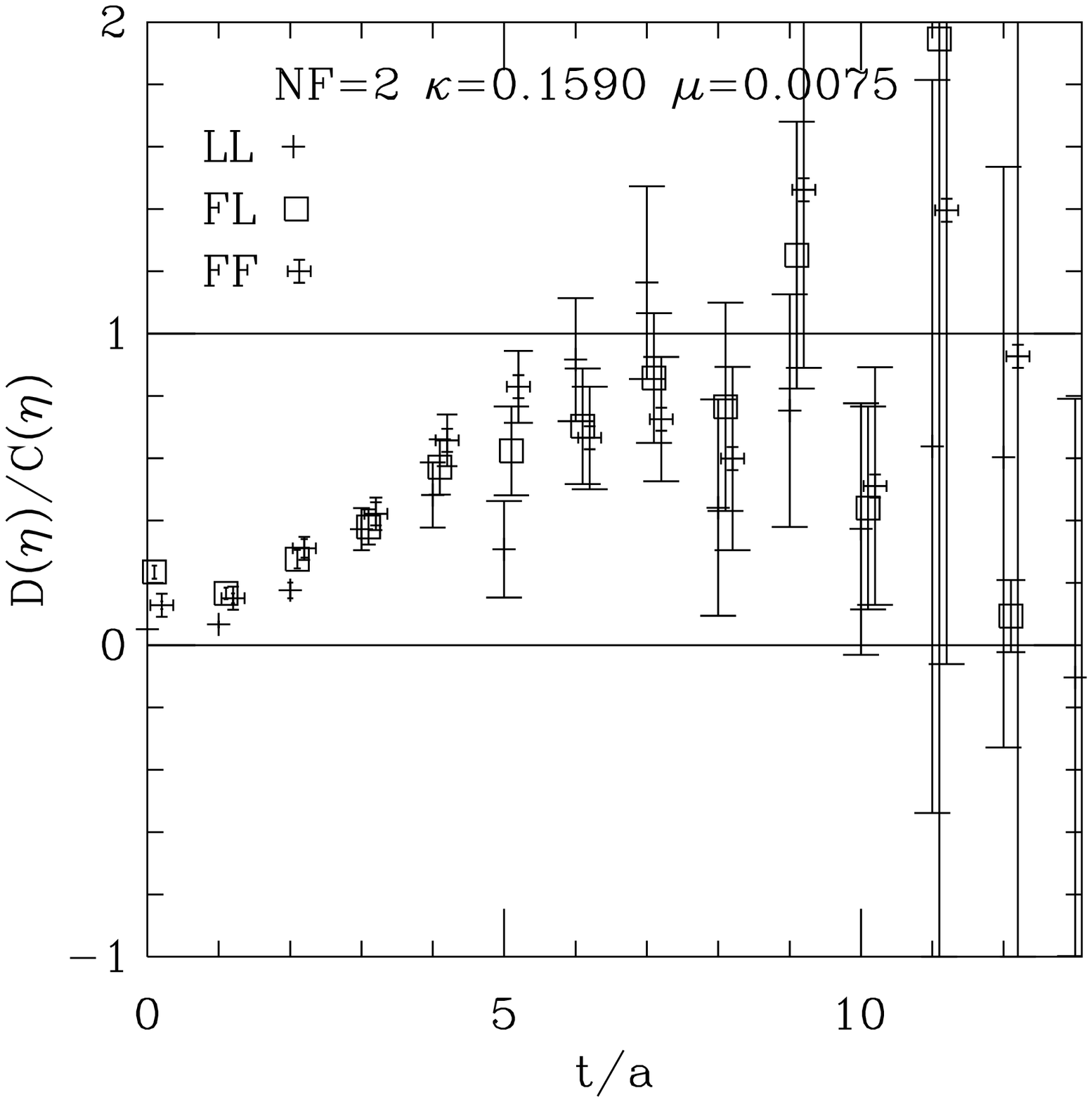}

 \caption{Flavour singlet pseudoscalar meson: ratio of disconnected
 to connected contribution to correlations. Here the total is $C-D$, 
so positivity implies $D/C < 1$. 
(a) quenched  and (b) with $N_f=2$.}

\end{figure}

We also explore the spectrum of scalar mesons. In non-twisted QCD,  the
flavour non-singlet $a_0$ receives only connected contributions while 
the singlet $f_0$ receives both connected and disconnected. In TMQCD, 
the $a_0$ states are again split with the $a_0^+$ having only connected 
contributions whereas $a_0^0$ has both. As shown in fig.~4, the
disconnected contribution  to $a_0^0$ does help to make it closer to the
charged $a_0$, although there is a residual flavour violation (that we
find decreases as the lattice  spacing decreases). 

 Again, because this time of the decay to $\eta \pi$, the quenched
correlators are unphysical (they are negative). The $N_f=2$ case,
however, shows that  positivity is restored, as it should be. 

 In detail, the scalar meson contributions will have to be corrected for
 the small admixture of pseudoscalar meson arising if the mixing angle
is not exactly $90^0$. This small admixture becomes significant at
larger $t$  because the pseudoscalar mass is lighter than the scalar
mass.

\begin{figure}[htb]

\vspace{6.5cm} 
\includegraphics{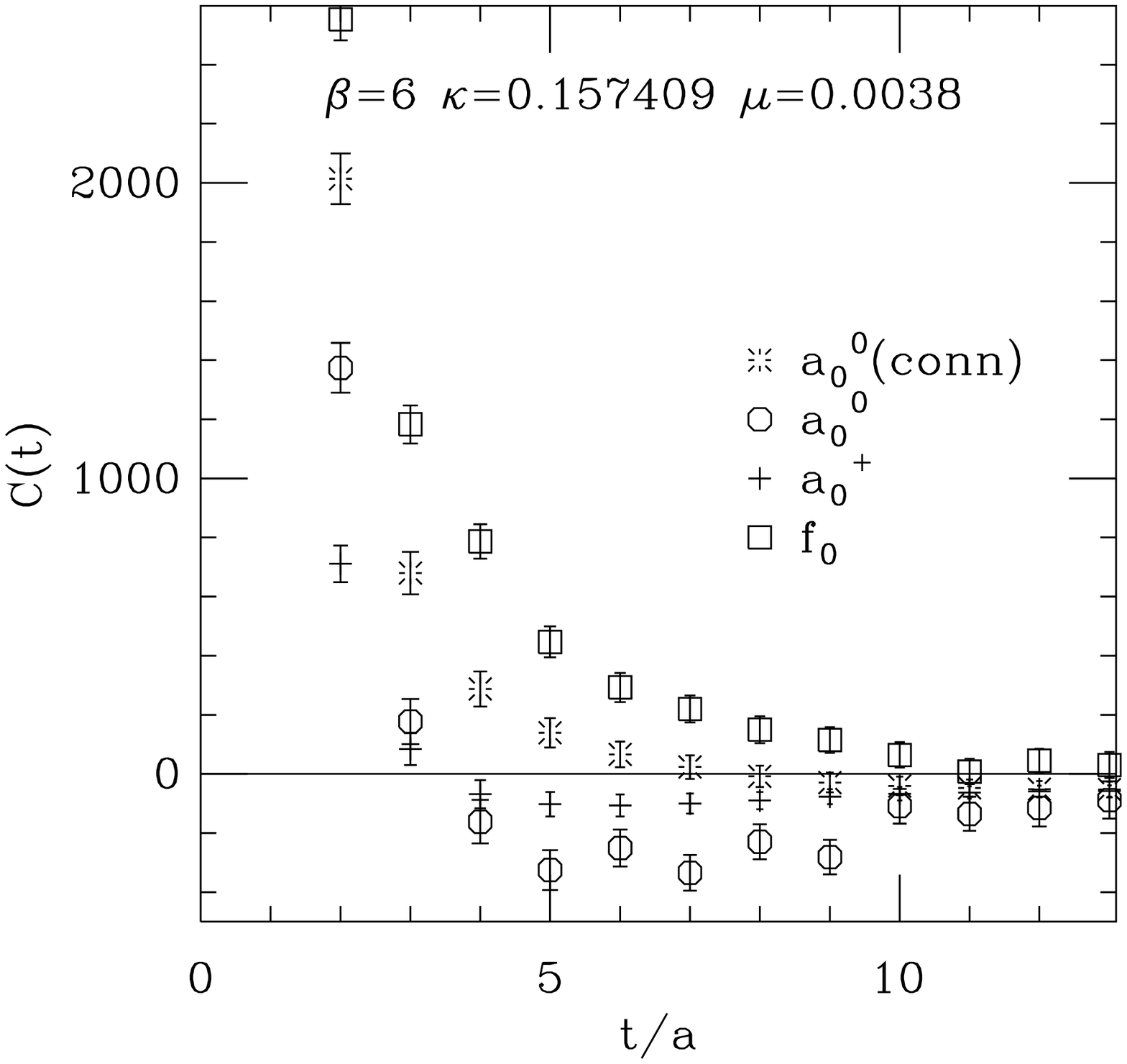}
\includegraphics{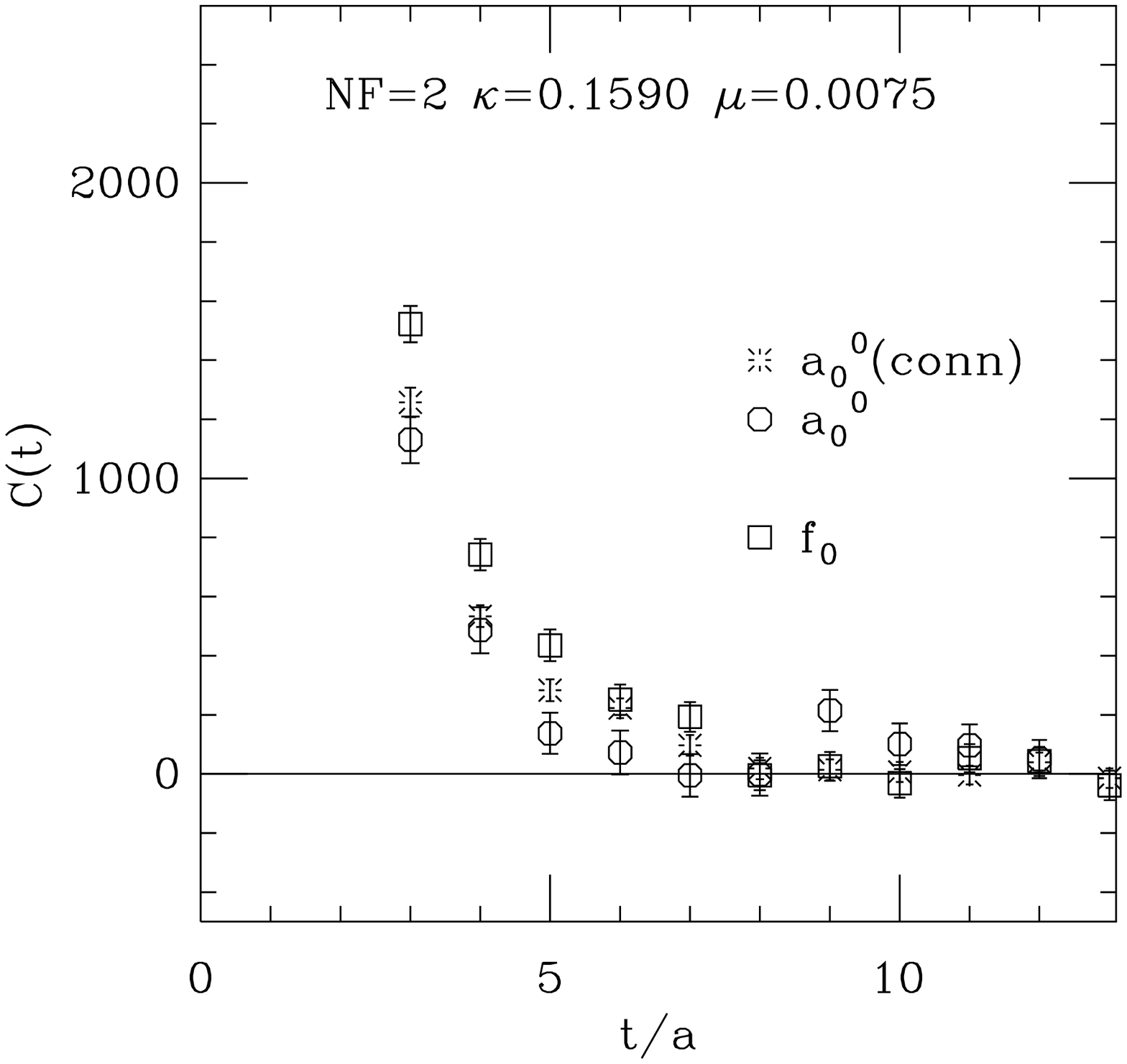}

 \caption{Scalar meson contributions (a) quenched and (b) with $N_f=2$}

\end{figure}

\section {Summary}

{ 
\begin{itemize}

 \item The disconnected contribution to $\pi^0$ is {\em big} compared to
the connected. 

 \item The $\pi^+ \ \pi^0$ mass difference decreases with decreasing $a$ as  
 $r_0^2 (m^2(\pi^0)-m^2(\pi^+)) \approx c(a/r_0)^2$
 in a quenched study with $c \approx 10$.

 \item From a  full-QCD study, one can extract the chiral  lagrangian
parameter relevant to $\pi^0$ $\pi^+$ splitting. For the  DBW2 gauge action,
the splitting is small (consistent with zero within errors)  with a 
sign which  corresponds to an Aoki phase at small quark mass. For the 
TLS gauge action, the same sign is also seen.

 \item The flavour singlet pseudoscalar meson signal ($\eta$) is sensible 
and useful for $N_f=2$ with dynamical TMQCD.

 \item For scalar mesons, the need for disconnected diagrams and the 
possibility of the signal being polluted by mixing (if twist not 90$^0$) 
with pseudoscalar suggests this will be a difficult area.

 \end{itemize}
}


\providecommand{\href}[2]{#2}\begingroup\raggedright\endgroup

\end{document}